\documentclass[aip,jap,amsmath,amssymb,reprint]{revtex4-1}
\usepackage{natbib}
\usepackage{graphicx}
\usepackage[utf8]{inputenc}
\usepackage{hyperref}

\begin{document}

\title{Physical characteristics and cation distribution of NiFe$_2$O$_4$ thin films with high resistivity prepared 
by reactive co-sputtering}

\author{C.\ Klewe}
\email{cklewe@physik.uni-bielefeld.de}
\author{M.\ Meinert}
\author{A.\ Boehnke}
\affiliation{Thin Films and Physics of Nanostructures, Department of Physics, Bielefeld University, 33501 Bielefeld, Germany}
\author{K.\ Kuepper}
\affiliation{Department of Physics, University of Osnabr\"uck, 49069 Osnabr\"uck, Germany}
\author{E.\ Arenholz}
\affiliation{Advanced Light Source, Lawrence Berkeley National Laboratory, CA 94720, USA}
\author{A.\ Gupta}
\affiliation{Center for Materials for Information Technology, University of Alabama, Tuscaloosa, Alabama, USA}
\author{J.-M.\ Schmalhorst}
\author{T.\ Kuschel}
\author{G.\ Reiss}
\affiliation{Thin Films and Physics of Nanostructures, Department of Physics, Bielefeld University, 33501 Bielefeld, Germany}

\date{\today}

\begin{abstract}
We fabricated NiFe$_2$O$_4$ thin films on MgAl$_2$O$_4$ (001) substrates by reactive dc magnetron co-sputtering in a pure oxygen atmosphere at different substrate temperatures. The film properties were investigated by various techniques with a focus on their structure, surface topography, magnetic characteristics, and transport properties. Structural analysis revealed a good crystallization with epitaxial growth and low roughness and a similar quality as in films grown by pulsed laser deposition. Electrical conductivity measurements showed high room temperature resistivity (12\,$\Omega$m), but low activation energy, indicating an extrinsic transport mechanism. A band gap of about 1.55\,eV was found by optical spectroscopy. Detailed x-ray spectroscopy studies confirmed the samples to be ferrimagnetic with fully compensated Fe moments. By comparison with multiplet calculations of the spectra we found that the cation valencies are to a large extent Ni$^{2+}$ and Fe$^{3+}$. 

\end{abstract}

\maketitle

\section{Introduction}
Recent advances in spintronics and spincaloritronics\cite{Bauer2012} have prompted the search for ferro- and ferrimagnetic insulating materials applicable in thin films.
Since the discovery of the spin Seebeck effect (SSE) in 2008 by Uchida \textit{et al.},\cite{Uchida2008} numerous publications focused on investigations of the SSE both in a transverse (TSSE) and in a longitudinal (LSSE) configuration.\cite{Uchida2010} However, in a longitudinal configuration the SSE can hardly be distinguished from the anomalous Nernst effect (ANE), if ferromagnetic, conducting materials are investigated.\cite{Meyer13} This makes observations of the LSSE in metallic thin films challenging.
More suitable materials for studies of the LSSE are ferromagnetic insulators.
The lack of free charge carriers in insulating materials like yttrium iron garnet (YIG) can prevent the appearance of ANE and, therefore, allows to identify the LSSE unequivocally.\cite{Uchida2010}
Also, a wide range of other spincaloric and spintronic effects, like the recently proposed spin Hall magnetoresistance (SMR)\cite{Nakayama2013, Chen2013} benefit from the supression of charge currents in magnetic systems with high resistivity. In SMR based devices, the insulating behaviour reduces shorting through the magnetic layer and can eliminate an anisotropic magnetoresistance (AMR). This enables to measure the SMR without parasitic effects.\cite{Althammer2013}

Some of the spinel ferrites are ferrimagnetic and semiconducting.
These materials are a promising alternative to several compounds of the garnet class\cite{Uchida2013} for a broad range of spincaloric and spintronic applications. 
For example, the LSSE was also observed in (Mn,Zn)Fe$_2$O$_4$,\cite{Uchida10} in nickelferrite (NiFe$_2$O$_4$)\cite{Meyer13} and in magnetite (Fe$_3$O$_4$) below the Verwey transition.\cite{Ramos13}
The class of spinel ferrites is characterized by the general formula AB$_2$O$_4$, where A and B denote divalent and trivalent cations, respectively.
Each conventional unit cell of a spinel ferrite consists of 8 formula units AB$_2$O$_4$, leading to 32 divalent oxygen anions forming a face centered cubic structure within the unit cell.
Two separate sublattices are formed, denoted as tetrahedral and octahedral, based on their coordination numbers 4 and 6 regarding the neighbouring oxygen anions, respectively. 

In our study we focus on the inverse spinel ferrite NiFe$_2$O$_4$ (NFO), which shows both LSSE\cite{Meyer13} and SMR.\cite{Althammer2013} 
In the inverse spinel structure half of the B$^{3+}$ cations reside on tetrahedral sites while the remaining B$^{3+}$ and A$^{2+}$ cations are located on octahedral sites.\cite{Verwey47}
NFO offers the advantage to switch on or off parasitic effects like the ANE by changing the temperature, due to its semiconducting properties. 
The compound is a ferrimagnet below the Curie temperature T$_\text{C}\approx850\,$K\cite{Pauthenet52, Dionne88}, with an antiferromagnetic coupling between the tetrahedral and octahedral sublattices.

A wide range of chemical and physical deposition techniques has been utilized to fabricate ferrite thin films. However, pulsed laser deposition (PLD) has been established as the most common method for synthesizing high quality ferrite thin films.\cite{Caltun04, Datta2010}
Here we report on the fabrication of high-quality epitaxial NFO thin films by reactive magnetron co-sputter deposition. 
This is beneficial for the integration of NFO in technological applications, as this method is compatible with a wide range of other fabrication techniques and allows for a high growth rate.

\section{Methods}
\subsection{Experimental methods}

Epitaxial NFO thin films were grown by ultra high vacuum reactive dc magnetron co-sputtering in a pure oxygen atmosphere at various substrate temperatures ($400^\circ$C to $800^\circ$C). 
The base pressure was lower than $10^{-8}\,$mbar. The O$_2$ pressure during sputtering was $2.2 \times 10^{-3}\,$mbar. The target-to-substrate distance was 21\,cm and the confocally arranged sources were tilted towards the substrate at an angle of $30^\circ$.
The films were co-sputtered from elemental Ni and Fe targets. 
All films were deposited on (001) oriented, isomorphous MgAl$_2$O$_4$ (MAO) substrates with a lattice mismatch of about $3\%$. 
The deposition rate was $0.14\,$\AA{}/s. The film thicknesses were 28\,nm for the temperature series and 58\,nm for samples deposited in a subsequent series at $680^\circ$C.

X-ray reflectivity (XRR) was used to calibrate the film thickness.
X-ray fluorescence (XRF) was used to quantify the Ni:Fe ratio.
In order to study the crystallographic properties of the films, x-ray diffraction (XRD) measurements were carried out in a Bragg Brentano configuration. 
The XRR, XRF, and XRD studies were performed in a Philips X'Pert Pro diffractometer with a Cu K$_\alpha$ source.

X-ray photoelectron spectroscopy (XPS) gave additional information about the cation valencies.
The XPS measurements were performed using a PHI5600ci multitechnique spectrometer equipped with a monochromatic Al K$_\alpha$ source (1486.7\,eV) with 0.3\,eV full width at half-maximum. The overall resolution of the spectrometer is 1.5\% of the pass energy of the analyzer, 0.45\,eV in the present case. Complementary Fe 2p spectra were also recorded with an Mg K$_\alpha$ standard non monochromatic x-ray source (1253.6\,eV). The measurements were recorded with the sample at room temperature. The thin NFO films were rinsed with Isopropanol just before mounting them into the loadlock of the experimental chamber. No other preparation of the sample surface, e.g. by Ar ion sputtering was performed, in particular to avoid a reduction of the Fe and Ni ions. The spectra were calibrated with a corresponding measurement of the Au 4$f_{7/2}$ level (84.0\,eV) of a gold foil.

The magnetic properties were investigated by alternating gradient magnetometry (AGM) in a Princeton MicroMag with magnetic fields of up to 13\,kOe. 

Atomic force microscopy (AFM) studies of the surface topography were done with a Bruker AFM Multimode instrument using Bruker FMV-A probes in tapping mode.

Optical spectroscopy in the range of $0.8$ to $5.5$\,eV was performed in a Perkin Elmer Lambda 950 Spectrometer. Reflection and transmission spectra were recorded to obtain the absorption coefficient and derive the optical bandgap.

The transport properties were investigated by temperature dependent dc conductivity measurements in a cryostat with a two-point probe technique. 

Element specific x-ray absorption (XAS), magnetic circular dichroism (XMCD), and magnetic linear dichroism (XMLD) measurements were taken at room temperature at beamline 4.0.2 of the Advanced Light Source, Berkeley. The substrate luminescence was detected with a photodiode to measure the absorption signal of the films in addition to the sample drain current (total electron yield). The magnetic field of $5000$\,Oe was switched for every energy, either in the film plane (XMLD) or parallel to the beam with the sample surface including an angle of 30$^\circ$ with the beam. The XMLD spectra were taken along the $[100]$ direction of the NFO film. The resolving power of the beamline was set to $E/\Delta E \approx 6000$. The degree of circular polarization was 90\%.

\subsection{Calculations}\label{sect:calc}

The absorption spectra were calculated within atomic multiplet and crystal field theory with the CTM4XAS program.\cite{CTM4XAS} The crystal field parameter for the octahedral Ni$^{2+}$ was set to $10Dq = 1.1$\,eV and the Slater integrals were reduced to 75\% and 90\% of their atomic values for the d-d and p-d interactions, respectively, to account for screening. For the octahedral (tetrahedral) Fe$^{3+}$ we chose $10Dq=1.6$\,eV ($10Dq = -0.8$\,eV). The Slater integrals were reduced to 75\% and 85\%. An exchange field of $g \mu_\mathrm{B} B = 10$\,meV was applied to break the spin symmetry. For Ni, a Lorentzian broadening of 0.15\,eV (0.3\,eV) was applied to the L$_3$ (L$_2$) edges to account for lifetime effects. For Fe, the Lorentzian broadening was set to 0.2\,eV (0.4\,eV). An additional Gaussian broadening of 0.15\,eV was applied to account for the finite resolving power of the instrumentation. Additionally, the calculated spectrum of the octahedral Fe$^{3+}$ was shifted by $0.23$\,eV to higher energy. Both species were weighted 1:1. These parameters were chosen to obtain a best fit to the experimental data and are close to the parameters suggested in earlier publications.\cite{Arenholz06, Arenholz07}

\section{Results and discussion}

\subsection{Variation of deposition temperature}

\subsubsection{Stochiometry and crystal structure}
By adjusting the sputter parameters according to the XRF observations, we obtained Ni:Fe ratios between 1.02:1.98 and 0.97:2.03 for the films, close to the correct composition. 
\begin{figure}[t]
	\centering
		\includegraphics{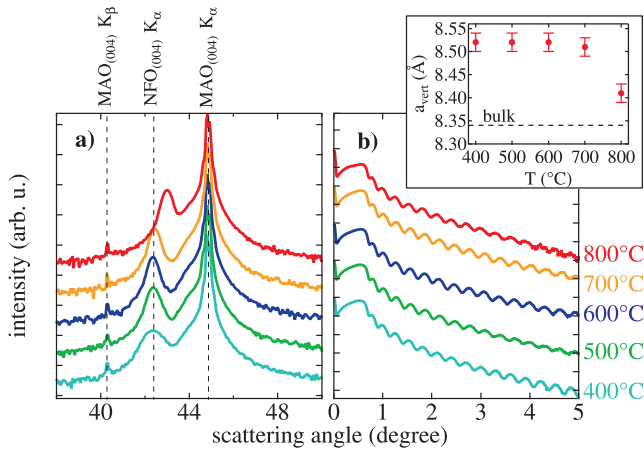}
	\caption{(Color online) a) XRD and b) XRR patterns of NFO films for different deposition temperatures. Inset: vertical lattice parameter a$_\mathrm{vert}$ plotted against deposition temperature.}	
	\label{fig:XRD}
\end{figure}
In FIG. \ref{fig:XRD}a) we present the results of XRD measurements. 
In the x-ray diffraction patterns (004) Bragg peaks are visible for all samples, i.e. all the films show a crystalline structure with epitaxial growth in $[001]$ direction. The peaks get more pronounced with increasing deposition temperature. Weak Laue oscillations were observed at the (004) peak in the 600$^\circ$C curve, indicating the smoothest interfaces for this deposition temperature.

A significant out-of-plane strain is visible in all samples. The deviations of the vertical lattice parameter with respect to the bulk lattice constant (a$_\mathrm{bulk}=8.34$\,\AA{}) can be derived from the peak positions in the XRD scans (see inset in FIG. \ref{fig:XRD}). 
The films deposited at 400$^\circ$C, 500$^\circ$C, and 600$^\circ$C show the same vertical lattice parameter a$_\mathrm{vert}=8.53$\,\AA{}, which is in good agreement with the values found by Foerster \textit{et al.} in pulsed laser deposited thin films on MAO, but is significantly larger than the bulk value.\cite{Foerster2011}
The increase of the vertical lattice parameter with respect to the bulk value implies that the films are tetragonally distorted, i.e. expanded in the direction perpendicular to the surface and compressed in the film plane, due to a comparatively large mismatch with the MAO(001) substrate (a$_\mathrm{MAO}=8.08$\,\AA{}).
However, with increasing deposition temperature the lattice parameter decreases leading to a$_\mathrm{vert}=8.51$\,\AA{} and a$_\mathrm{vert}=8.41$\,\AA{} for the films sputtered at 700$^\circ$C and 800$^\circ$C, respectively. 
This trend in lattice distortion can be explained by an increased strain relaxation at higher thermal energies.

The well pronounced oscillations in the x-ray reflectivities shown in FIG. \ref{fig:XRD}b) indicate smooth surfaces for all sputtered films with roughnesses in the range of 0.3\,nm to 0.4\,nm. The films deposited at 600$^\circ$C and 700$^\circ$C showed the smallest values while the 800$^\circ$C sample showed the highest roughness.

\subsubsection{Magnetic properties}

\begin{figure}[t]
	\centering
		\includegraphics{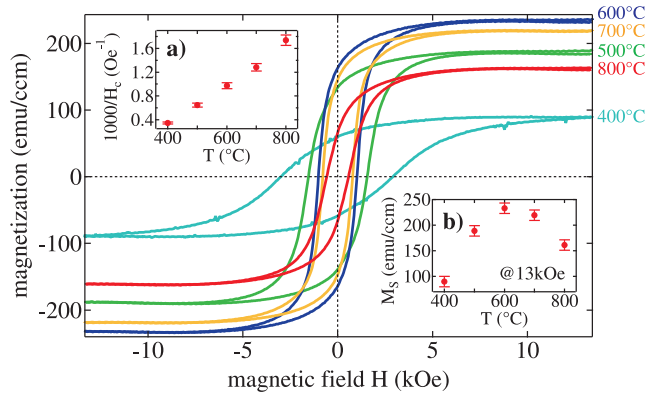}
	\caption{(Color online) Magnetization curves (AGM)of NFO for different deposition temperatures. Inset a): The reciprocal coercivity H$_\mathrm{C}^{-1}$ over deposition temperature. Inset b): Saturation magnetization at 13\,kOe over deposition temperature.}	
	\label{fig:AGM}
\end{figure}
FIG. \ref{fig:AGM} displays the magnetization after subtraction of a linear background. 
It is observed that the coercivity decreases with increasing temperature (see inset a)). 
Since a lower coercivity can be generated by less pinning centers and, therefore, less defects in the material, this reciprocal dependence between coercivity and deposition temperature confirms a lower defect density for higher deposition temperatures.\cite{Kuschel2012} 

The saturation magnetization shows a different behaviour. It increases from 400$^\circ$C to 600$^\circ$C, but decreases slightly for 700$^\circ$C and drastically for 800$^\circ$C deposition temperature. 
This trend in the temperature dependence is depicted in inset b) of FIG. \ref{fig:AGM}. The highest saturation magnetization of about 236\,emu/cm$^3$ is observed for deposition at 600$^\circ$C. Since there seems to be no generally accepted value for the bulk saturation magnetization we can only state that it probably lies between 270\,emu/cm$^3$ and 300\,emu/cm$^3$.\cite{Venzke96, Rigato07, Lueders06} 
A lower magnetization in thin films with respect to bulk has been observed in NFO before,\cite{Venzke96} as well as in other inverse spinel ferrites like magnetite (Fe$_3$O$_4$).\cite{Moussey04} The reduced magnetization at 13\,kOe may arise from antiferromagnetic pinning in antiphase boundaries, the density of which critically depends on the preparation conditions.\cite{Bobo01, Datta2010} Another mechanism of reduction of the magnetization may be the formation of disordered regions between crystallites.\cite{Venzke96} The large coercive field of the films (around $1000$\,Oe) is in agreement with a relatively high defect density. In both cases, full saturation is obtained at much higher fields, so the background subtraction in measurements with only a few Tesla may be incorrect.

\subsection{Sample deposition at 680$^\circ$C}

\subsubsection{Structural and magnetic properties}
Based on the former results, additional samples were produced at a substrate temperature of 680$^\circ$C with a thickness of about 58\,nm and more detailed studies were carried out. 
The samples again crystallized well with a (001) orientation and showed smooth surfaces in the XRR diffraction patterns. 
The vertical lattice parameter a$_\mathrm{vert}=8.48$\,\AA{} is less expanded than in the corresponding films from the temperature series, due to an increased lattice relaxation with increasing thickness. We also derived an in-plane lattice parameter a$_\mathrm{ip}=8.19$\,\AA{} from an analysis of the (066) peak. As expected, the values reveal a pronounced tetragonal distortion for growth on MAO (001) substrates with $a_\mathrm{vert}/a_\mathrm{ip} = 1.035$. The unit cell volume is reduced by about 2\% with respect to bulk material.
The additional samples had a slightly lower magnetization (around 200 emu/cm$^3$) than the films prepared earlier for the deposition temperature series. No out-of-plane anisotropy was observed, in line with the theoretical prediction by Fritsch {\it et al.}\cite{Fritsch2010} for NFO films with in-plane compressive strain.

\begin{figure}[t]
	\centering
		\includegraphics{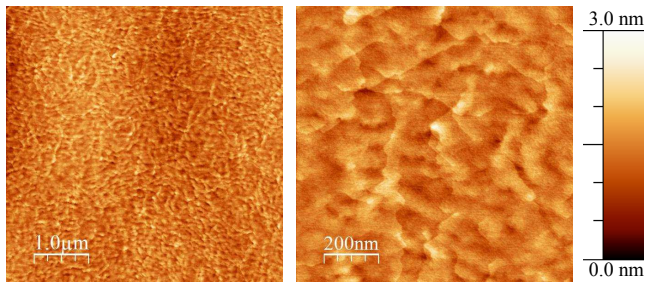}
\caption{(Color online) AFM measurements on a 58\,nm NFO thin film sputtered at 680$^\circ$C. Left: $5\,\mu \mathrm{m} \times5\,\mu \mathrm{m}$ scan range. Right: $1\,\mu \mathrm{m} \times1\,\mu \mathrm{m}$ scan range. In both scan ranges the roughness is remarkably low (R$_\text{RMS}$=0.27\,nm, R$_\text{avg}$=0.21\,nm).}	
	\label{fig:AFM}
\end{figure}
To confirm the low roughness, additional AFM measurements were conducted.
Images of a $5\,\mu \mathrm{m} \times5\,\mu \mathrm{m}$ and a $1\,\mu \mathrm{m} \times1\,\mu \mathrm{m}$ tapping mode scan are shown in FIG. \ref{fig:AFM} revealing a smooth and homogeneous growth on the MAO (001) substrate.
In both cases the roughness is about 0.27\,nm (RMS) and 0.21\,nm (average).
The small roughness of the films is beneficial for the integration of NFO in spincaloric devices, as it enables high interface quality and, therefore, high spin mixing conductances across the interfaces. This might increase the effect amplitude for spin caloric effects considerably.\cite{Qui2013}

The following sections refer to studies on the 58\,nm thick films grown at 680$^\circ$C.

\subsubsection{Electronic properties}
\begin{figure}[t]
	\centering
		\includegraphics{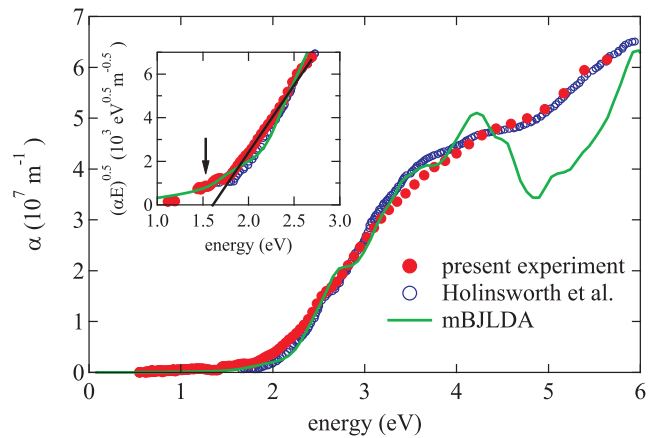}
\caption{(Color online) Optical absorption spectrum taken at RT for a film sputtered at 680$^\circ$C. For comparison, experimental data from Ref. \onlinecite{Holinsworth2013} and a theoretical calculation from Ref. \onlinecite{Meinert13} are shown. Inset: Tauc plot $(\alpha\mathrm{E})^{0.5}$ versus energy for the determination of the minimum gap. The arrow indicates the mBJLDA minimum direct gap.}	
	\label{fig:UV/VIS}
\end{figure}

FIG. \ref{fig:UV/VIS} depicts the optical absorption spectrum of an NFO film in an energy range from 0.8\,eV to 5.6\,eV. Experimental data by Holinsworth {\it et al.}\cite{Holinsworth2013, Sun2012} and a calculated spectrum by Meinert {\it et al.}\cite{Meinert13} are shown for comparison. 
The absorption coefficient $\alpha(\mathrm{E})$ was obtained from the measured transmission and reflectance spectra by $\alpha = \frac{1}{d} \mathrm{ln}  \left( \frac{1-R}{T} \right)$. The absorption spectrum of our NFO sample is very similar to both that of the PLD fabricated thin film (Holinsworth et al.) as well as to the calculated spectrum. 
The Tauc plot ($(\alpha\mathrm{E})^{0.5}$ versus energy) displayed in the inset of Figure \ref{fig:UV/VIS} helps to determine the minimum gap, which seems to be slightly smaller than for the PLD film, as indicated by the straight line in the inset. From this plot, we obtain $E_\mathrm{gap} \approx 1.55$\,eV.  In contrast to common notion, it has been shown that this type of plot does not necessarily indicate the presence of a direct gap in NFO.\cite{Meinert13}

\begin{figure}[b]
	\centering
		\includegraphics{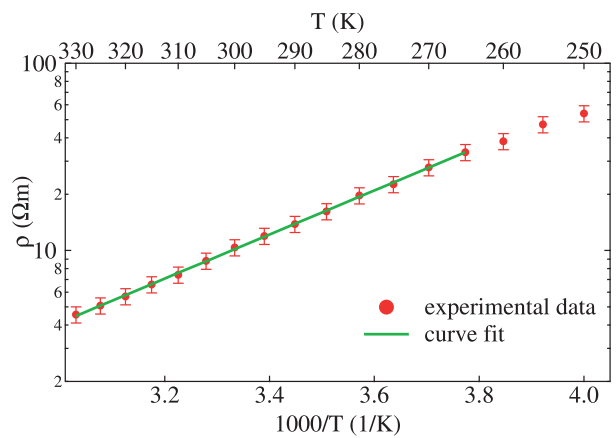}
	\caption{(Color online) Temperature resolved resistivity measurement on a sample sputtered at 680$^\circ$C. From the curve fitting an activation energy of 0.23\,eV was derived. The error in the data is mainly arising from uncertainty in the contacting and does not significantly alter the outcome of the curve fitting.}
	\label{fig:conductivity}
\end{figure}

Conductivity measurements were performed to study the electronic transport properties.
A sample was patterned into a strip of 950\,$\mu\mathrm{m}$ length and 70\,$\mu\mathrm{m}$ width.
The results are shown in FIG. \ref{fig:conductivity}.
At 295\,K we find a RT resistivity of $\rho \approx 12\,\Omega \mathrm{m}$, corresponding to a measured resistance of about 3\,G$\Omega$. For the resistivity we estimate an uncertainty of at least 15\%, mainly arising from locally not well-defined contacting. 
This resistivity is one to three orders of magnitude larger than values found in other sputtered NFO thin films, which were deposited in a mixed Ar/O$_2$ atmosphere.\cite{Lueders06, Jin2010}

The required energy for the thermally activated charge transport can be derived by a linear regression of the temperature dependent resistivity.
The $\rho$ vs. 1/T curve was fitted with the equation $\rho \propto \mathrm{exp}({\frac{E_\mathrm{a}^\mathrm{ext}}{k_{\mathrm{B}}T}})$ for impurity induced conduction in extrinsic semiconductors in order to determine the thermal activation energy. In the high temperature region above 265\,K a straight line segment in the ln($\rho$) plot is present and can be fitted with the above relation (see FIG. \ref{fig:conductivity}). 
For the specimen investigated in this work we obtain an activation energy E$_\mathrm{a}^\mathrm{ext}=0.23$\,eV. This result is of the same order as the values found by Lord \textit{et al.} in sintered NFO specimens, Austin \textit{et al.} in NFO single crystals, and Ponpandian \textit{et al.} in NFO nanoparticles.\cite{Lord60, Austin70, Ponpandian2002} 
However, the value is six times smaller than the bandgap we find by optical spectroscopy. 
Since electrical conductivity is sensitive to all charge transport mechanisms present in the film, chemical impurities can significantly lower the observed activation energy.

It has been suggested that the electric transport in ferrites is driven by charge carrier exchange between divalent and trivalent cations on equivalent lattice sites.\cite{Verwey41, vanUitert55, Lord60} 
In the case of NFO, small amounts of excess Fe or Ni might enter the lattice on octahedral sites during the preparation process as Fe$^{2+}$ or Ni$^{3+}$ ions, respectively.\cite{Lord60, Austin70} Alternatively, an oxygen deficit would lead to incomplete oxidation of Fe atoms, generating Fe$^{2+}$ instead of trivalent species.
This promotes $\mathrm{Fe}^{2+} \leftrightarrow \mathrm{Fe}^{3+}$ or $\mathrm{Ni}^{2+} \leftrightarrow \mathrm{Ni}^{3+}$ hopping processes of electrons or holes, respectively.\cite{vanUitert55, Verwey50} 
De Boer \textit{\textit{et al.}} proposed that such electron interchange processes require only little energy, as the charge carriers travel along the statistical cation distribution on the octahedral sites without altering the energy state of the lattice considerably.\cite{deBoer37}
According to the very high resistivity of our films, we conclude that only a small number of such defects is present and that the stoichiometry is close to the desired NFO composition. In the next section we investigate this aspect in more detail.

\subsubsection{Cation distribution and element resolved magnetic moments}
\begin{figure}[t]
\includegraphics[width=8.5cm]{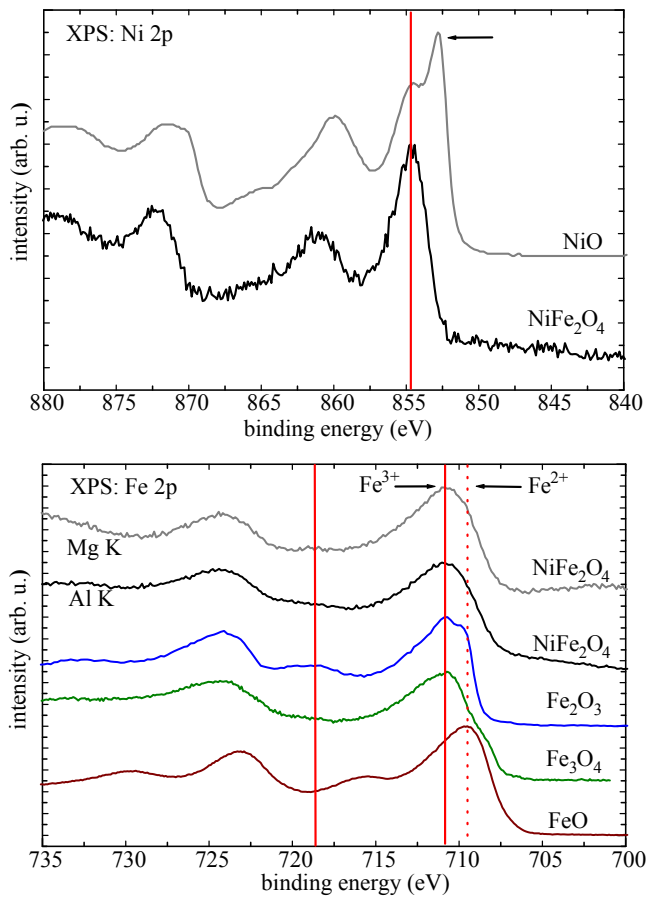}
\caption{\label{fig:XPS}Upper panel:  The Ni 2p XPS of NFO with a corresponding NiO spectrum\cite{uhl92} as a reference. Lower panel: the Fe 2p spectrum of NFO recorded with Al K$\alpha$ and Mg K$\alpha$ source, respectively. Fe 2p spectra of $\alpha$-Fe$_2$O$_3$, Fe$_3$O$_4$, and FeO for comparison\cite{pri05, bru13} are also shown.}
\end{figure}

Ni 2p and Fe 2p core level XPS spectra were taken to evaluate the valencies of Ni and Fe.
The Ni 2p XPS spectrum of NFO  (FIG. \ref{fig:XPS} (upper panel)) shows the Ni 2p$_{3/2}$ (854.2\,eV) and Ni 2p$_{1/2}$ (871.6\,eV) main peaks followed by a rich satellite structure due to corresponding charge transfer excitations. Except an intense peak at the low binding energy side of NiO \cite{uhl92} (marked by an arrow), which can be attributed to an intrinsic feature found for NiO \cite{vee93, tag08}, the Ni 2p spectra of NFO and NiO are very similar to each other, which is consistent for Ni$^{2+}$ ions in a high spin state. 
The lower panel of FIG. \ref{fig:XPS} displays the Fe 2p XPS of NFO along with reference spectra of 
$\alpha$-Fe$_2$O$_3$, Fe$_3$O$_4$, and FeO for comparison.\cite{pri05, bru13} The Fe 2p$_{3/2}$ binding energy of 710.8\,eV matches that of the corresponding Fe$_2$O$_3$ peak. However, since the Fe 2p spectrum recorded with the monochromatic Al K$_\alpha$ source is to some extent overlapped by an  Ni L$_3$M$_{23}$M$_{45}$ Auger peak the typical charge transfer feature for Fe$^{3+}$ ions in oxidic materials (around 719\,eV) appears to be obscured. Therefore, we performed an additional measurement with a standard non monochromatic Mg K$_\alpha$ source. Here the Ni L$_3$M$_{23}$M$_{45}$ Auger does not overlap with the Fe 2p spectrum and also the characteristic charge transfer feature appears, even though somewhat weaker in intensity compared to that of Fe$_2$O$_3$. However, the Fe 2p spectrum recorded with the Mg K$_\alpha$ source resembles the hard x-ray PES spectrum of  Jaffari \textit{et al.}\cite{Jaffari2012}. Hence, we observe a divalent Ni and a trivalent Fe valence state in this NFO thin film.

\begin{figure}[t]
\includegraphics[width=8.5cm]{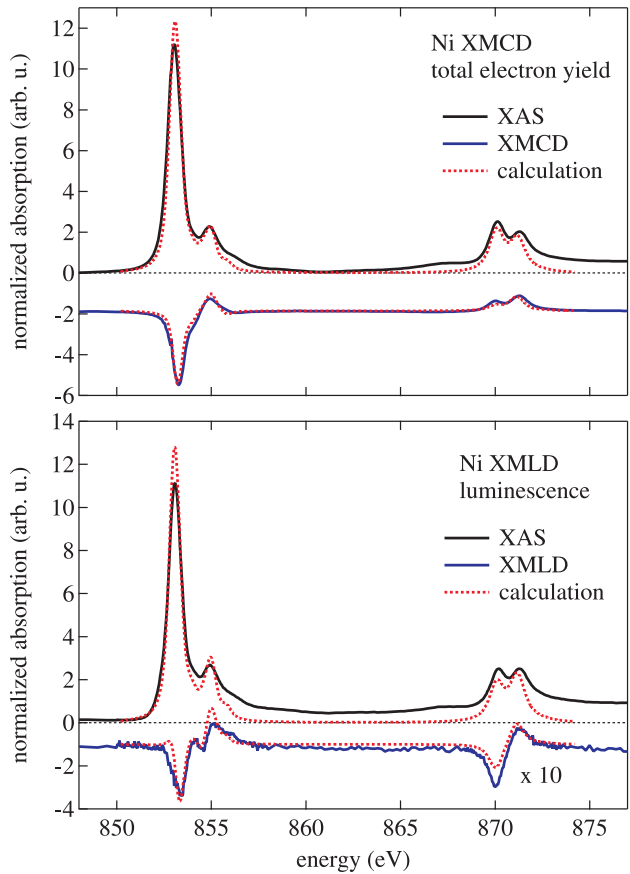}
\caption{\label{fig:Ni_xray}Experimental and computed XAS, XMCD, and XMLD spectra of Ni in NFO. The XAS spectrum is normalized to 1 at 40 eV above the L$_3$ onset.}
\end{figure}

\begin{figure}[t]
\includegraphics[width=8.5cm]{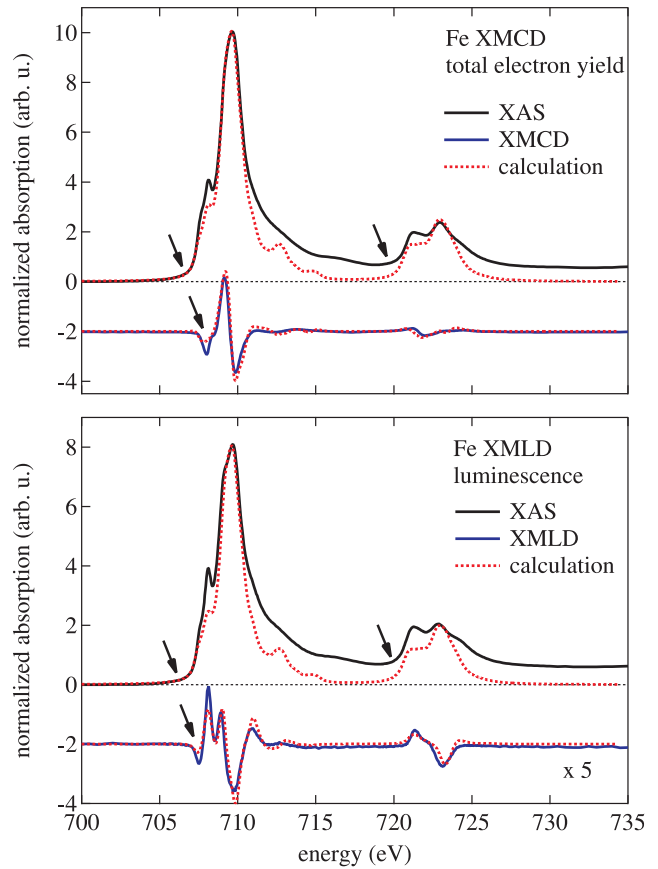}
\caption{\label{fig:Fe_xray}Experimental and computed XAS, XMCD, and XMLD spectra of Fe in NFO. The arrows mark positions at which characteristic features of Fe$^{2+}$ should appear if it was present on octahedral sites. The XAS spectrum is normalized to 1 at 40 eV above the L$_3$ onset.}
\end{figure}

Experimental and computed XAS, XMCD, and XMLD spectra of Ni and Fe are shown in Figures \ref{fig:Ni_xray} and \ref{fig:Fe_xray}. The agreement between experimental and computed spectra for Ni is almost perfect, indicating that Ni$^{2+}$ occupies solely octahedral sites. 
Also for Fe a good fit to our measurements is obtained with the parameters given in section \ref{sect:calc}.
In particular, no characteristic features of Fe$^{2+}$ in the octahedral sites (as in Fe$_3$O$_4$) are found, such as a shoulder at the onset of the L$_3$ and L$_2$ XAS spectra, a much larger negative first peak in the L$_3$ XMCD spectrum or a positive peak at the onset of the L$_3$ XMLD spectrum.\cite{Arenholz06,Lu04} The corresponding positions are marked by arrows in FIG. \ref{fig:Fe_xray}. Thus, it may be concluded that there is only a small fraction Fe$^{2+}$ present on the octahedral positions, in agreement with the XPS results.

Evidence for the complete structural inversion (Fe occupies equally tetrahedral and octahedral positions) comes from the sum rule analysis of the Fe XMCD spectrum. The total spin magnetic moment is $m_\mathrm{spin}^\mathrm{Fe} = (0.10 \pm 0.05)\,\mu_\mathrm{B}$ per atom, where the rather large relative uncertainty comes from the non-trivial absorption background structure. The ratio of orbital and spin magnetic moments is $m_\mathrm{orb}^\mathrm{Fe}/m_\mathrm{spin}^\mathrm{Fe} = (0.12 \pm 0.02)$. Thus, although the XMCD signal of Fe is large in amplitude, pointing to the presence of large magnetic moments, they compensate nearly completely. This is exactly what is expected for the inverse spinel structure, in which the tetrahedral and octahedral Fe$^{3+}$ sites give rise to different XMCD spectra, so they do not compensate. However, the magnetic moments should cancel nearly exactly.\cite{Szotek06,Fritsch2010} In the normal spinel structure, Fe$^{3+}$ would only occupy octahedral sites and have parallel magnetic moments. Consequently, mainly the Ni$^{2+}$ sites contribute to the macroscopic magnetization. The sum rule analysis of the Ni spectra is difficult due to the complicated background. However, the orbital to spin moment ratio does not depend on the normalization, so it may be deduced with good accuracy. We find $m_\mathrm{orb}^\mathrm{Ni}/m_\mathrm{spin}^\mathrm{Ni} = (0.24 \pm 0.02)$, in agreement with an earlier estimate.\cite{vanderLaan99} With the total magnetization of $m = 200\,\mathrm{emu}/\mathrm{cm}^3$, i.e. $1.53\,\mu_\mathrm{B}$\,/\,f.u. from the AGM measurement, we deduce the spin magnetic moment of Ni to be $m_\mathrm{spin}^\mathrm{Ni} = (1.06 \pm 0.15)\,\mu_\mathrm{B}$ and the orbital magnetic moment $m_\mathrm{orb}^\mathrm{Ni} = (0.25 \pm 0.10)\,\mu_\mathrm{B}$.

\section{Conclusion}
NFO thin films were fabricated by reactive dc magnetron co-sputter deposition in a pure oxygen atmosphere.
X-ray diffraction studies revealed that the films crystallized well with low roughnesses on MAO (001) substrates at different deposition temperatures.
A mismatch induced strain was visible for all films, which relaxes for higher deposition temperatures and thicknesses.
Magnetic investigations reflected the ferrimagnetic behaviour of the NFO thin films with a reduced magnetization compared to bulk samples. 
From the obtained data the ideal deposition temperature for device oriented application was derived to be between 600$^\circ$C and 700$^\circ$C.
An additional series sputtered at 680$^\circ$C with comparable crystallographic and magnetic properties showed a low roughness and bandgaps very similar to values found in films deposited by pulsed laser deposition.
XPS, XAS, XMCD, and XMLD spectra yield a cation distribution as expected for an inverse spinel structure. A nearly complete compensation of the Fe moments on tetrahedral and octahedral sites was observed. The macroscopic magnetization is mainly carried by the Ni$^{2+}$ ions.
A semiconducting behaviour with a low activation energy of E$_\mathrm{a}^\mathrm{ext} \approx 0.23\,$eV and a large RT resistivity of about $\rho \approx 12\,\Omega$m was confirmed by conductivity measurements, further promoting the utilization of sputtered NFO thin films in spincaloric and spintronic applications.

\section*{Acknowledgments}
The authors gratefully acknowledge financial support by the Deutsche Forschungsgemeinschaft (DFG) within the Priority Program \textit{Spin Caloric Transport} (RE1052/24-1). K.K. acknowledges further support by the DFG (KU2321/2-1). They thank D. Meier for helpful discussion and U. Heinzmann for making available the UV/Vis spectrometer.
They are grateful for the opportunity to work at BL 4.0.2 of the Advanced Light Source, Berkeley, USA, which is supported by the Director, Office of Science, Office of Basic Energy Sciences, of the US Department of Energy under Contract No. DE-AC02-05CH11231.

\end{document}